\documentclass[12pt]{article}
\input{epsf.sty}
\usepackage{hyperref}
\usepackage{amsmath, amssymb}
\newtheorem {thm}{Theorem}[section]

\newtheorem {lem}[thm]{Lemma}

\newcommand{\qed}{\nobreak \ifvmode \relax \else
      \ifdim\lastskip<1.5em \hskip-\lastskip
      \hskip1.5em plus0em minus0.5em \fi \nobreak
      \vrule height0.75em width0.5em depth0.25em\fi}

\def\Cox{\hfill \Box}

\def\Z{{\Bbb Z}}

\def\P{{\Bbb P}}

\def\S{{\Bbb S}}

\def\ba{{\backslash}}
\def\sb{{\subset}}

\def\a{\alpha}

\def\ba{\setminus}
\def\b{\beta}

\def\d{\delta}

\def\phi{\varphi}

\def\g{\gamma}

\def\l{\lambda}

\def\r{\rho}

\def\s{\sigma}

\def\t{\tau}

\def\z{\zeta}

\def\L{\Lambda}

\def\O{{\Omega}}

\def\P{{\Phi}}

\def\T{\T}

\def\GG{{\cal G}}

\begin{document}
\title{Discrete approximations  to vector spin   models}

\author{
Aernout C.D. van Enter
\footnote{ University of Groningen,Johann Bernoulli Institute of Mathematics and
Computing Science, Postbus 407, 9700 AK Groningen, The
Netherlands,
\newline
 \texttt{A.C.D.v.Enter@math.rug.nl}, 
\newline
\texttt{http://www.math.rug.nl/~aenter/ }}
 \, ,  Christof K\"ulske
\footnote{ Ruhr-Universit\"at   Bochum,Fakult\"at f\"ur Mathematik, D44801 Bochum, Germany,
\newline
\texttt{Christof.Kuelske@ruhr-uni-bochum.de}, 
\newline
\texttt{http://www.ruhr-uni-bochum.de/ffm/Lehrsttuehle/Kuelske/kuelske.html
/$\sim$kuelske/ }}\, 
\, \\ and
 Alex A. Opoku 
\footnote{ Mathematisch Instituut, Universiteit Leiden, 
\newline
Postbus 9512, 2300 RA, Leiden , the Netherlands,
\newline
\texttt \texttt{opoku@math.leidenuniv.nl}
\newline
 \texttt{
http://www.math.leidenuniv.nl/nl/staff/223/ }}
}

\maketitle

\begin{abstract} 
We strengthen a result from \cite{KO} on the existence of effective 
interactions for discretised continuous-spin models. We also point out that 
such an interaction cannot exist at very low temperatures.  
Moreover, we compare two ways of discretising continuous-spin models, and show 
that, except for very low temperatures, they behave similarly in two 
dimensions. We also discuss some possibilities in higher dimensions.

 \end{abstract}



\vfill\eject

\section{Introduction}
If one tries to approximate a continuous-spin vector model such as the 
classical XY-model by a discrete approximation, whether for 
computational or for theoretical purposes \cite{Kac}, 
one can in principle pursue two routes. 
\begin{enumerate}
\item Either one could consider the Gibbs 
measure for the original continuous-spin model, and discretise the spin 
by dividing the single-spin space into a large but finite number $q$ of sets 
(intervals).
By identifying all spins in such a set, one obtains a measure on a 
discrete-spin system. This measure can, but does not have to, be a Gibbs measure 
for some effective discrete-spin interaction.

\item Or, alternatively, one could write down the same expression for the 
Hamiltonian of  the discrete-spin system as one has for the XY-model, and 
consider the appropriate Gibbs measure(s) for this discrete spin interaction.  
For the XY-model (the plane rotor) such models are called clock models, and 
their study goes back to Potts \cite{Po}.
\end{enumerate}

Here we present some results on discrete approximations of type 1, and 
compare them with what is known on discrete approximations of type 2. We will 
see that, except for very low temperatures, in two dimensions both 
approximations have quite similar properties, and we speculate on possible 
scenarios also in higher dimensions. 

 We notice that a type 1 discretisation is appropriate for measuring 
purposes, and describes for example round-off errors, whereas a type 2 
discretisation is what often happens in computer simulations. 

Generalisations to discretisations of more general continuous compact 
single-spin spaces are immediate, as follows from the analysis of \cite{KO,O}. 

 Stated differently, and more formally, we can apply 
a local discretisation map $T:\S^1 \mapsto \{1,\dots, q\}$, mapping 
a continuous local spin variable, taking values on the circle $\S^1$, to its discretized image, before 
or after performing the Gibbsian modification with interaction  $\Phi$ 
relative to the product measure $\a$. 

We then want to 
compare the images $T \mu$ of the Gibbs measures $\mu\in \GG_{\Phi,\a}$ 
of the initial model with a priori measure $\a$ and interaction  $\Phi$ 
with the Gibbs measures $\mu'\in \GG_{\Phi,T\a}$ where $T \a$ is the product 
of the a priori measures under the local coarse-graining $T$ 
and $\GG_{\Phi,T\a}$ are the Gibbs measures obtained from the specification 
which has the old interaction simply taken in the coarse-grained variables. 

\smallskip 

We note, by the way, that conceptually such a question can be studied even 
more generally without making any assumptions on the first and/or second 
image-spin measure being discrete.

A first important question to be asked is whether and when 
$$T\GG_{\Phi,\a} = \GG_{\Phi',T\a}$$  
for some $\Phi'$. 

We remind the reader that if there is an $\Phi'$ such that 
$T \mu \in \GG_{\Phi',T\a}$ for a $\mu \in \GG_{\Phi,\a}$, then 
$T\GG_{\Phi,\a} \sb \GG_{\Phi',T\a}$ \cite{EFS}.  

We are not aware that equality  between the number of transformed 
Gibbs measures and the number of Gibbs measures for the transformed 
interaction -even if one exists- always holds, although we don't know of
any counterexamples. We also note that, even under the assumption of  equality  
of these two sets, one can have a different number of extremal Gibbs measures 
in the original and the transformed set,  if different measures 
are mapped to the same one, as e.g. occurs in Fuzzy Potts models.

We can also ask questions of closeness 
on the level of interactions, namely, what is the distance between original and 
transformed interaction $d(\Phi,\Phi')$?  
Furthermore, what is a good notion for the distance $d$ here? 


There is the problem here 
that the spins,  and hence the interactions,  live on different spaces, 
one discrete, one continuous. 
If one compares the two discretisations one has  at least the advantage that 
the corresponding interactions will  live on the same space.  

What can be said about closeness of the measures 
$\mu\in \GG_{\Phi,\a}$ to $\tilde \mu\in \GG_{\Phi,T\a}$? 
This question is subtle, since 
we could look here for closeness on local observables,--that is, 
in the weak topology-,  
on the level of long-range characteristics like decay of correlations, 
on the level of the phase diagram in parameter space, ... 
Here we  will call the two discretisations close,  if the two models have 
a similar phase diagram and{/} or similar correlation decay in 
their Gibbs measures.

We  note that the proof concerning the locality properties  in \cite{KO} for 
type-1 discretisations makes essential use of the Dobrushin
uniqueness theorem, even though we need not be in the uniqueness regime and 
in fact are allowed to be in a phase-transition region when we discretise. 
This will also be the case here.

Discretisations can be viewed as single-site coarse-grainings, similarly to the 
fuzzification or amalgamation of discrete-spin systems as treated in e.g. 
\cite{CU,H,HK,Ver}, but now the ``fuzzification'' goes from a continuous ``alphabet" to a discrete one. 
\section{Gibbsianness of discrete approximations of the XY  lattice model}

\subsection{Notation and Definitions}
We will consider lattice spin systems with a single-spin space $\Omega_{0}$, on a 
lattice $\Z^d$, and a configuration space $\Omega= {\Omega_{0}}^{\Z^d}$.
We will mainly consider the XY-model, for which $\Omega_0$ is the circle $\S^1$, and 
discrete approximations thereof, in which  $\S^1$ is divided into $q$  
equal arcs of length $\frac{2 \pi}{q}$.  We will indicate the spin 
variables at site $i$ (which always will be elements of the unit circle) by 
$\sigma_i$, $\omega_i$, $\eta_i$, and 
similarly spin configurations in a volume $\Lambda$ 
by $\sigma_{\Lambda}$, $\omega_{\Lambda}$, $\eta_{\Lambda}$.
 
We will consider Gibbs measures, which are defined for 
(here translation-invariant) absolutely summable 
interactions $\Phi$ (that is,  $\sum_{A;\; 0 \in A} ||\Phi_{A}|| < \infty$) via the DLR equations, expressing that given an 
external configuration $\eta_{\Lambda^c}$, 
the probability density of configurations in 
a volume $\Lambda$ is given by the Gibbs expression 
\begin{equation}\label{inimod}
\begin{split}
\frac{d\mu_{\Lambda}^{\eta_{\Lambda^c}}}{d\alpha_\Lambda}(\sigma_{\Lambda}) &= 
\frac{\exp \left(-  H_{\Lambda}(\sigma_{\Lambda}\eta_{\Lambda^c})\right)}{Z_{\Lambda}^{\eta_{\Lambda^c}}},\quad \text{where}\quad
H_{\Lambda} (\s_\L\eta_{\L^c})=\sum_{A;\; A\cap \L\neq \emptyset}\beta \P_A(\s_\L\eta_{\L^c}),
\end{split}
\end{equation}
and $\alpha_{\Lambda}$ is the product of $\alpha$ over the sites in $\Lambda$.
This should hold for all volumes $\Lambda$, internal configurations $\sigma_{\Lambda}$ and external configurations $\eta_{\Lambda^c}$.
The corresponding collection of 
(everywhere instead of almost everywhere with respect to the Gibbs measure 
defined) conditional probabilities forms a ``specification''.
See e.g. \cite{EFS,DEZ,Geo}.  
In the standard nearest-neighbour models, (the plane rotor or XY-model), as well as in
 the clock models, where the spins take discrete values, we have 
\begin{equation}
 -H_{\Lambda}(\s_\L\eta_{\L^c}) = \beta \sum_{< i,j > \in \Lambda}\sigma_i \cdot \sigma_j + \beta \sum_{< i \in \Lambda, j \in \Lambda^c >} \sigma_i \cdot \eta_j .
\end{equation}


\subsection{Conservation of Gibbsianness under local transformations: fine discretisations}


One of the main results of \cite{KO}, see also \cite{O}, concerns conditions under which a 
discretisation of a Gibbs measure is again Gibbsian. These results were obtained as corollaries 
to a theorem on the preservation of Gibbsianness which also holds for much more general types 
of local transforms like time evolutions. So, it is worthwhile to reconsider specifically the local transformations.

The condition for preservation of Gibbsianness is temperature-dependent, and the main example we want to discuss here is the discretisation of 
the circle into $q$ equal arcs.  At inverse temperature $\beta$ the result 
implies that for $q$, dependent on 
$\beta$, large enough, the discretised measure is a Gibbs measure. 

To be more precise, suppose for each $l\in S':=\{1,2,\cdots,q\}$ we denote by $\S^1_l$ the $l$th 
arc of the circle $\S^1$ cut out by the discretisation operator $T$. Then, $T\alpha(l)=\alpha(\S^1_l)$.
Given $\mu\in \mathcal{G}_{\Phi,\alpha}$, one of the main results in \cite{KO}, Theorem 2.5, is that the discretised  measure $T\mu$  is Gibbs if 
\begin{equation}\label{Dobest}
\sup_{i\in\Z^d}\sum_{j\in \Z^d\setminus\{i\}}\bar C_{ij}<1,
\end{equation}
where 
\begin{equation}
\begin{split}
2\,\bar C_{ij}=\left\{\begin{array}{ll}\sup_{\eta_j,\bar\eta_j\in\S^1;\atop l\in S'}\int_{\S_l^1}\alpha(d\sigma_i)
\left|\frac{e^{\beta\,\sigma_i\cdot\eta_j}}{\int_{\S_l^1}\alpha(d\hat\sigma_i)\,
e^{\beta\,\hat\sigma_i\cdot\eta_j}}-\frac{e^{\beta\,\sigma_i\cdot\bar\eta_j}}{\int_{\S_l^1}\alpha(d\hat\sigma_i)\,
e^{\beta\,\hat\sigma_i\cdot\bar\eta_j}}\right|, &\mbox{ if } |i-j|=1,\\\\
0, &\mbox{otherwise.}
\end{array}
\right.
\end{split}
\end{equation}
Thus, constrained on a  discrete-spin configuration, the constrained system 
must be  in the Dobrushin uniqueness  regime {\em uniformly in the chosen 
constraint}. 

While looking for good upper bounds for the right-hand side, we can at not 
much additional cost 
revisit the more general situation and give an improvement to the criterion 
from \cite{KO} for Gibbsianness for local discretisations.


We put ourselves in a 
 slightly more general context than that of 
the discretisations in  \cite{KO}, and we will take the local spin space $S$ 
just to be a general compact measurable space. 
No a priori metric is given; it will be produced by the Hamiltonian itself. 
As in \cite{KO}, let a decomposition be given of the form $S=\bigcup_{s'\in S'}S_{s'}$. Here 
$S'$ may be a finite or infinite set. 
Put $T(s):=s'$ for $S_{s'}\ni s$. This defines a deterministic 
transformation on $S$, called the {\it fuzzy map} (the discretisation).  

Now we deviate from \cite{KO}.
Let $G$ be the vertex set of a general graph and define a family of metrics $ (d_{ij})_{j\in G\ba \{i\}}$ on 
the local spin space at the site $i\in G$
by 
\begin{equation}\begin{split}\label{lippi}
& d_{ij}(\s_i,\tau_i):=\sup_{{\zeta,\bar \zeta}\atop{\zeta_{j^c}=\bar \zeta_{j^c}; T(\zeta_{j})=T(\bar\zeta_{j})}} \Bigl
| H_i(\s_i\z_{i^c})-H_i(\s_i\bar\z_{i^c})-\Big(H_i(\tau_i\z_{i^c})-
H_i(\tau_i\bar\z_{i^c})\Big)\Bigr |, 
\end{split}
\end{equation}
where  for any $i\in G$,  $i^c=G\setminus \{i\}$.
 It is important here (as well as in the formula specific to the rotors above) 
that the supremum  is taken over spins $\zeta, \bar \zeta$ which are 
constrained to take the same coarse-grained image at $j$. We are allowed to do 
this since we are analyzing the constrained system. In this way the metric at 
the site $i$ depends also on the size of the coarse-graining at $j$. 
The metric measures how strongly  a variation at the site $j$ can maximally 
change the difference in interaction energy between local spins $\s_i,\tau_i$.

Our criterion of 
the fineness
of the decomposition will involve the corresponding {\it $j$-diameter}, namely the quantity  
$\text{diam}_{ij}(A)=\sup_{s,t \in A}d_{ij}(s,t)$ 
where $A$ 
runs over the  sets in the decomposition.

\begin{thm}\label{fuzzy} 
Let $\mu$ be a Gibbs measure of the specification with Gibbsian potential 
$\Phi$ with an arbitrary a priori measure $\a$, on a graph with vertex set $G$.
Let $T$ denote the local coarse-graining map where we assume that 
$\a(S_{s'})>0$ for all labels $s' \in S'$.    

Suppose that   
\begin{eqnarray}
\sup_{i\in G}\sum_{j\in G \ba i }\sup_{s'}\text{diam}_{ij}(S_{s'})< 4.
\end{eqnarray}

Then the transformed measure $T(\mu)$ is Gibbs for a specification $\g'$ with 
an absolutely summable discrete-spin interaction $\Phi'$. 

\end{thm}

In all cases this is an improvement over the criterion of \cite{KO} 
(which we don't repeat here in detail, 
because it  requires the introduction of additional structure which we don't need here.) 

 It is also an improvement over what a direct application of the high-temperature 
version found in Georgii \cite{Geo}  would give for our constrained model. That would  only give a bound in terms of the right-hand side of 
the inequality of the form
\begin{equation}\begin{split}\label{lippilies}
& \text{diam}_{ij}(S_{s'})\leq 2  \sup_{\s_i,\t_i,\zeta,\bar \zeta} \Bigl
| H_i(\s_i\z_{i^c})-H_i(\t_i\bar\z_{i^c})\Bigr|.
\end{split}
\end{equation}

Here we discuss the application to rotor models. 
Consider first the rotor model on a circle $\S^1$. 
We have for n.n. $i$ and $j$ by Cauchy-Schwartz that 
\begin{equation}\label{flippi}\begin{split}
 d_{ij}(\s_i,\t_i)&=
\b \sup_{\zeta_j,\bar \zeta_{j}; T(\zeta_{j})=T(\bar\zeta_{j})} \Bigl|(\s_i-\t_i)\cdot (\zeta_j-\bar \zeta_j)\Bigl| \cr
&\leq \beta\Vert \s_i - \t_i \Vert_2  2\sin \frac{\pi}{q}
\end{split}
\end{equation}
and so $\text{diam}_{ij} S_{s'} = \b \times (2 \sin \frac{\pi}{q})^2$. 
This gives the criterion 
\begin{equation}
 2 d \b  (\sin \frac{\pi}{q})^2 < 1
\end{equation} 
 for Gibbsianness of the coarse-grained model. 
 Note that the standard estimate \eqref{lippilies} would give a worse condition without the square. 

For a local spin space which is a $d$-dimensional sphere not much changes. 
The formula for the metric $d_{ij}$ stays the same.   
Let us assume that $\psi$ is one half of the maximal angle under which 
a set $S_{s'}$ appears as seen from the origin. This quantity is a measure of fineness of the discretisation.   
Then, going through the same steps, 
we obtain as a criterion for Gibbsianness that 
\begin{equation}
 2 d \b  (\sin \psi)^2 < 1.
\end{equation}

\bigskip 

\noindent
\textbf{Proof of the Theorem:} 
The proof follows as in \cite{KO}, by  estimating $\bar C_{ij}$. This constant is a bound on the Dobrushin interaction matrix 
of the initial model conditional on the transformed spins, uniformly in the values of the transformed spins. In particular, for each   
 site $i\in G$,  $\bar C_{ij}$ is a uniform upper bound on    the variational distance between  the ``first-layer models'' in $\{i\}$, in which $\sigma_i$ is 
constrained to take  values in $ S_{s'}$ for 
some fixed prescription of partitions given by the image spins $\s'_i$, wrt   external configurations that coincide everywhere except at site $j$. 

%
%
%


More precisely, we take two conditioning configurations in the original (first-layer) model, 
$\z,\eta\in \O$ with $\z_{j^c}=\eta_{j^c}$  and denote by
$u_0(\s_i)=-H_i(\s_i\zeta_{i^c})$ and
$u_1(\s_i)=-H_i(\s_i\eta_{i^c})$ the corresponding values 
of the single-site Hamiltonians anchored at $i$. 
Defining $u_t=t u_1+(1-t)u_0$,   
   $h_t=e^{u_t}1_{S_{\s'_i}}/\a(e^{u_t}1_{S_{\s'_i}})$ and
$\l_t(d\s_i)=h_t(\s_i)\a(d\s_i)$, with $t\in[0,1]$, we note that
$\l_0(d\s_i)=\g_i(d\s_i|\z)$ and
 $\l_1(d\s_i)=\g_i(d\s_i|\eta)$, where $\g_i$'s are the single-site parts of the conditional 
 distributions (kernels) of the initial model obtained via \eqref{inimod} after replacing $S$ with $S_{\s'_i}$.  Note however that 
the constraining configuration  $\s'$ does not appear any more 
 in the notation, for the sake of simplicity. 

Now comes the estimate which  for coarse-grainings 
improves the one from \cite{KO}
(in which, however,  also general transformations beyond coarse-grainings were 
treated), and also \eqref{lippilies} from Georgii. 
 For the first step of the proof, we obtain the following bound 
for the total variational norm of $\l_0-\l_1 $;
 \begin{equation}\label{linear7}
 \begin{split}
& 2\Vert\l_0-\l_1 \Vert =\int \a(d\s_i) |h_1(\s_i)-h_0(\s_i)|= \int
 \a(d\s_i) \Bigl |\int_{0}^1 dt\frac{d}{dt}h_t(\s_i)\Bigr |\\
&\leq \int_0^1 dt\;
\int\l_t(d\s_i)\Bigr|H_i(\s_i\z_{i^c})-H_i(\s_i\eta_{i^c})-\int\l_t(d\tau_i)\Big(H_i(\tau_i\z_{i^c})-H_i(\tau_i\eta_{i^c})\Big)\Bigl|.
\\
\end{split}
\end{equation}
The simple but essential next 
estimate will be uniform in the image measure of $\l_t$, under $\s_i\mapsto H_i(\s_i\z_{i^c})-H_i(\s_i\eta_{i^c})=:f^{\z,\eta}(\s_i)$.  
Namely, by further making use of the notion of the $j$-diameter of the set $S_{\s'_i}$ to bound 
the variation of the energy terms we get     
\begin{equation}\label{linear}
 \begin{split}
&\int\l_t(d\s_i)\Bigr|H_i(\s_i\z_{i^c})-H_i(\s_i\eta_{i^c})-\int\l_t(d\tau_i)\Big(H_i(\tau_i\z_{i^c})-H_i(\tau_i\eta_{i^c})\Big)\Bigl| \cr
&\leq \int\l_t(d\s_i)\int\l_t(d\tau_i)|  f^{\z,\eta}(\s_i)-f^{\z,\eta}(\tau_i)       | \\
&= \int f^{\z,\eta}(\l_t)(d x)\int f^{\z,\eta}(\l_t)(d y)| x-y     | \\
&\leq \sup_{\l } \int_{- D}^D \l(dx) \int_{- D}^D dy \l(dy)|x-y| \\
\end{split}
\end{equation}
where $D=\text{diam}_ {i j} \, S_{\s'_i}/2$ and the supremum is over the probability measures $\l$ on the interval $[-D,D]$. 
For this supremum we use the following lemma (after scaling with $D$).

\begin{lem} For all probability measures $\r$ on $[-1,1]$ we have \\
$Q(\r):=\int\r(dx) \int \r(dy) |x-y|\leq 1$ with equality for $\r_0=\frac{1}{2}(\d_1+\d_{-1})$. 
\end{lem}

Note the improvement over the simple upper bound $2$. Observe also that the upper bound on $2\|\l_0-\l_1\|$ \eqref{linear7}  obtained via 
\eqref{linear} is independent of $\eta, \zeta\in\Omega$ and constraint $\sigma_i'$.
Therefore \eqref{linear} provides a uniform  upper bound on $2 \bar C_{ij}$. By
scaling up the interval in the Lemma  with a factor $D$ and putting together our previous estimates 
we get $2 \bar C_{ij}\leq    \sup_{s'}\text{diam}_{ i j} \, S_{\s'_i}/2$, and hence $\sum_{j: j\neq i} \bar C_{ij}\leq \frac{1}{4}
\sum_{j\in G \ba i }\sup_{s'}\text{diam}_{ij}(S_{s'})$. The rest of the proof follows from the definition of the Dobrushin constant.
This proves the Theorem. 
$\Cox$
\bigskip 
\bigskip 

For the sake of completeness let us also give an elementary proof of the Lemma.

\bigskip

\noindent {\bf Proof of the Lemma.}
By density arguments we can approximate any $\r$ by convex combinations of finitely 
many Dirac measures of the form $\sum_{i=1}^n p_i \d_{x_i}$ where $x_i \leq x_{i+1}$. 

Let us look at $Q$ as a function of the $l$-th location, keeping the other locations fixed,  and keeping the $p_i$'s fixed, 
$x_l \mapsto Q(\sum_{i=1}^n p_i \d_{x_i})$, where $x_l$ is constrained to be greater than or equal to its left neighbor $x_{l-1}$ 
and less than or equal to its right neighbor $x_{l+1}$. This function is linear.
Hence the function takes its maximum when $x_{l}$ becomes 
equal to one of its neighbors. This shows that the maximum of $Q$ over the set of combinations 
of $n$ Dirac measures is dominated by that over combinations of $n-1$ Dirac measures. Iterating this argument we see that 
the maximum of $Q$ over all probability measures 
is reached for a linear combination of two Dirac measures $p \d_x +  (1-p) \d_y$. Noting finally that the max over 
$Q (p \d_x +  (1-p) \d_y)= 2 p(1-p)|x-y| $ is reached for $\r_0$ we are done. 
$\Cox$

 As far as the bound on $\sum_{j\in\Z^{d}\setminus\{i\}}\bar C_{ij}$ is concerned, the above result is an improvement over Theorem 2.9 of \cite{KO}
 which however was formulated in a much more general situation. 
 Indeed,  the latter gave rise to the bound $\sum_{j\in\Z^{d}\setminus\{i\}}\bar C_{ij}\leq \frac{4\,d\,\pi\, \beta}{q}\,e^{\beta}.$
The more general set-up of \cite{KO} allowed also to treat 
(partially) stochastic single-site maps, such as infinite-temperature 
stochastic dynamics.  The estimates  on the  Dobrushin constant used there 
were  of the ``high-field'' type , whereas here we make use of a 
``high-temperature'' version.

Once the refinement is large enough ($q$ very large at a fixed temperature)
the effective interaction has as its dominant term the 
nearest-neighbor interaction of the clock model. 

We notice that the discretised model inherits various properties 
from the original XY-model.  
In particular, if the correlation functions decay slowly, as they do in 
two dimensions at low temperature when one is in a Kosterlitz-Thouless phase
this remains true after the discretisation. 

The continuous symmetry of the original model is also inherited.  
In three and more dimensions there is a continuum of Gibbs measures, 
as the circle symmetry of the original XY-model is broken. 
Discretising the spin space results again in a continuum of Gibbs measures 
(which now however are not related to a broken symmetry of the discrete-spin 
model).

 We have thus proved the following theorem. 

\begin{thm} For each $d\geq 3$ there is a $q_0$ such that for $q \geq q_0$  
there is an interaction $\Phi'$ 
with a discrete -clock - rotation invariance 
such that 
there are uncountably many translation-invariant ergodic states in the set of 
Gibbs measures 
$\GG_{\Phi' }$ (taken with uniform a-priori measures).  
\end{thm}

 This argument provides an independent rigorous route to 
the existence of an intermediate enhanced-symmetry 
Kosterlitz-Thouless phase in a  discrete-spin model, combining 
our general criteria for preservation of Gibbsianness under local 
coarse-grainings with properties  of the original continuous-spin model.

\section{Comparing the discretisations}

At high temperatures, in the paramagnetic regime, everything is well-behaved, 
but not of great physical interest. We will therefore discuss what happens in 
subcritical-temperature regimes.

It is a remarkable fact that the standard nearest-neighbor 
large-$q$ clock model in two dimensions has the property that there is a 
Kosterlitz-Thouless phase with slow decay and an enhanced continuous symmetry 
occurs at an intermediate temperature regime \cite{FS1, FS2}. On symmetry enhancement, see also \cite{NS82}.

The values for which this occurs are such that $q$ should be large enough, 
for a given low temperature. 
As we have just seen, the discretised XY-model can be described by a summable 
interaction --in which the nearest-neighbor terms are the dominant ones--, 
in just such an intermediate regime.

On the other hand, the nearest-neighbor 
clock model at fixed $q$ and at {\em very} low temperatures ($\beta \geq 
O(q^2)$) will have $q$ ordered 
phases, that is $q$ different Gibbs measures, similar to the $q$ ground states,  
all with exponential correlation decay. 
This follows directly from a Pirogov-Sinai argument. On the contrary,
we can  show that for a fixed even $q$ once the temperature is low enough 
(how low depends on $q$), the -type 1- discretised Gibbs measure becomes 
non-Gibbsian.
Indeed, if we take an alternating configuration for the discretised spin this 
implies  that alternatingly the spin is either in the most Northern (on sites in
one sublattice) or the 
most Southern interval (on the other sublattice) of size $\frac{2 \pi}{q}$. 
We argue that such a configuration is a point of 
essential discontinuity for a conditional probability  of the discretised 
measure.  
Conditioned on this, the original spins (which are forced by the constraint on 
which we condition to be almost opposite, but by their interaction prefer to be 
pointing in the same direction), will have two ground states, one pointing 
alternatingly North-West, South-West, and  the other one alternatingly 
North-East and South-East. 
The deviations in the Western , cq Eastern, direction are of order 
$O(\frac{1}{q})$, which means that the energy gap between the two ground states
 is of order $O(\frac{1}{q^2})$. Therefore, at sufficiently low temperatures ($ \beta \geq O(q^2))$, there will be two 
different Gibbs measures for the constrained model, and this will imply the 
non-Gibbsianness of the discretised measure. 
The details of the  argument  can be worked out in a straightforward manner  
along the lines of \cite{ekor, ER,Ruth}, see also \cite{Craw}.



Thus the analogy between the two discretisations breaks down just in this very-low-temperature regime.
The measures then are not even close any more on the level of local observables.
Since in this regime one finds very different behaviour, one discretisation 
resulting in a non-Gibbsian measure, 
and the other one in $q$ different Gibbs measures.

In higher dimensions, for the XY-model there is a continuum of Gibbs measures 
\cite{FSS} at low temperatures, which, as indicated above, are mapped to a 
continuum of different Gibbs measures for the discrete 
spins in an intermediate regime.


It would be interesting to see if the 
restoration of continuous symmetries which happens for the two-dimensional 
clock model would have a higher-dimensional analogue, in that in some 
intermediate-temperature regime there might exist a continuum of Gibbs measures,
 even for the nearest-neighbor clock model. We conjecture that the intermediate 
phase studied in \cite{ HWS,Osh,TUM,SI93,Ue,UK}, might be of this type. In the 
terminology of Ueno et al \cite{UK} we would have a continuum of 
``Incompletely Ordered Phases'', where the order can be in the two 
spin directions $n,n+1$, where $n \ mod \ q \in \{1,\cdots,q\}$,  
with continuously varying weights of these directions. 
Although there seems to be some doubt whether there exists  an intermediate 
phase at all in a region in between the $q$ ordered, ground-state-like, 
phases and the high-temperature paramagnetic phase, 
the numerical results up to now for 
the nearest-neighbor clock model appear to be inconclusive.  
It therefore seems worthwhile to investigate if an
``enhanced--broken-symmetry phase'' as decribed above, which can be 
obtained by discretizing a continuous-spin model,
could also occur for the nearest-neighbor clock model. 

The breakdown of the analogy at very low temperatures holds for the same reason 
as in two dimensions. The only property we used was the bipartiteness of the 
lattice. If we choose the North in the direction of the magnetisation, 
the arguments are unchanged. 


As we have just seen the transition between the Gibbsian behaviour 
and the non-Gibbsian behaviour occurs at  $\beta \geq O(q^2)$ for 
type-1 discretisations.

The  analysis of \cite{FS1} for type-2 discretisations similarly appears 
to provide  a transition value at $\beta= O(q^2)$ for the transition. 
For some numerical results, indicating this asymptotics in more detail, see
 \cite{TO}.

A heuristic reason for this behavior is that the model in the scaled variables $q$ 
times spin-angles 
with discretisation width $1$ approximates a discrete Gaussian model at 
effective inverse temperature $\b/q^2$; if this parameter is below the value 
for the roughening transition (which is rigorously known to take place in the 
discrete Gaussian) the model behaves like a massless Gaussian, while above 
it behaves like a massive model in the Peierls regime. 
Compare also Theorem C on page 40 of \cite{FS3}. 
There it is mentioned as a conjecture that, at fixed $q\geq 5$, the threshold 
values in temperature between low temperature regime and intermediate regime 
on the one hand, and intermediate and high temperature regime on the other 
hand, should be sharp, and different. See also \cite{BRP2010} for 
some numerical support for this.

\section{Conclusions}

We showed how to compare two different ways of discretising spin models, 
namely either starting from the Gibbs measures, for which we have a controlled 
approximation in a temperature-dependent regime, or starting from the 
interactions. We extended the regime in which we have such a controlled 
approximation of the discretised Gibbs measure, and also pointed out that it
cannot be extended to very low temperatures.  Thus the results are 
essentially optimal.  
In the two-dimensional XY-model both discretisations display 
the same Kosterlitz-Thouless phase in an  intermediate-temperature regime.

As for the higher-dimensional case, we  suggested the possibility of an  
enhanced-continuous-symmetry-breaking 
phase occurring at a region of intermediate temperatures in discrete 
$d$-dimensional clock models for $d$ at least three.



\begin{thebibliography}{99}


\bibitem{BRP2010} A.~F. Brito, J.A.~Redinz and J.A.~Plascak: Two-dimensional XY and clock models studied via the dynamics generated by rough surfaces, Phys. Rev. E 81, 031130 (2010).



\bibitem{CU} J.-R.~Chazottes and E.~Ugalde: On the preservation of 
Gibbsianness under symbol amalgamation. 
In: Entropy of hidden Markov processes 
and connections to Dynamical Systems. Eds. B.~Marcus, K.~Petersen and 
T.~Weissman. LMS Lecture Notes 385, to appear (2011).
arXiv 0907.0528. 






\bibitem{Craw} N.~Crawford: On Random Field Induced Ordering in the Classical 
XY Model, J. Stat. Phys. 142, 11-42 (2011).

\bibitem{EFS} A.C.D.~van Enter, R.~Fern\'andez, A.D.~Sokal: Regularity properties 
and pathologies of position-space renormalization-group transformations: Scope and 
limitations of Gibbsian theory, J. Stat. Phys. 72, 879-1167 (1993).

\bibitem{ekor} A.C.D.~van Enter, C.~K\"ulske, A.A.~Opoku and W.M.~Ruszel: Gibbs-non-Gibbs properties for $n$-vector lattice and mean-field models, 
Braz. J. Prob. Stat. 24, 226-255 (2010).







\bibitem{ER} A.C.D.~van Enter, W.M.~Ruszel: Gibbsianness vs. Non-Gibbsianness of
 time-evolved planar rotor models,   Stoch. Proc. Appl. 119, 1866--1888 
(2009).

 
 






\bibitem{DEZ} R.~Fern\'andez: Gibbsianness and non-Gibbsianness in lattice random 
fields, Les Houches, LXXXIII, (2005).

\bibitem{FS1} J.~Fr\"ohlich and T. Spencer: 
The Kosterlitz-Thouless transition in two-dimensional Abelian spin systems 
and the Coulomb gas, Comm. Math.Phys. 81, 527--602 (1981).


\bibitem{FS2} J.~Fr\"ohlich and T. Spencer: Massless phases and symmetry restoration in Abelian Gauge symmetries and spin systems, Comm. Math. Phys. 83, 411--454 (1982).

\bibitem{FS3} J.~Fr\"ohlich and T. Spencer: 
The Berezinskii-Kosterlitz-Thouless transition. In ``Scaling and Self-Similarity
in Physics'', J.Fr\"ohlich (ed.), Progress in Physics, Birkh\"auser, 
Basel and Boston (1983). 



\bibitem{FSS}
J.~Fr\"ohlich, B.~Simon, T.~Spencer:
Infrared bounds, phase transitions and continuous symmetry breaking,  
Comm. Math. Phys. 50, 79-95 (1976).

\bibitem{Geo} H.-O.~Georgii: Gibbs measures and phase transitions, volume 9 of de Gruyter Studies
in Mathematics. Walter de Gruyter Co., Berlin, 1988. ISBN 0-89925-462-4 (1988).







\bibitem{H} O.~H\"aggstr\"om: Is the fuzzy Potts model Gibbsian?
Ann. de l'Institut Henri Poincar\'e (B) Prob. and Stat. 39, 891-917 (2003).

\bibitem{HK} O.~H\"aggstr\"om, C.~K\"ulske: Gibbs properties of the fuzzy Potts model on
 trees and in mean field, Markov Proc. Rel. Fields 10 No. 3, 477-506 (2004).


\bibitem{HWS} R.K.~Heilmann, J.S.~Wang and R.B.~Swendsen: Rotationally 
symmetric ordered phase in the three-state antiferromagnetic Potts model, 
Phys. Rev. B 53, 2210 (1996).


\bibitem{Kac} We remind the reader of Mark Kac' famous dictum:  `` Be wise, discretise!''. 









\bibitem{KO} C.~K\"ulske, A.A.~Opoku: The Posterior metric and the
 Goodness of Gibbsianness for transforms of Gibbs measures, 
Electron. J. Probab. 1307--1344 (2008).












%



\bibitem{NS82} C.M.~Newman, L.S.~Schulman: Asymptotic symmetry: Enhancement and stability, Phys. Rev. B 26, 3910--3914 (1982).

\bibitem{O}
A.A.~Opoku: On Gibbs properties of transforms of lattice and mean-field systems,
Groningen thesis (2009).

\bibitem{Osh} M.~Oshikawa:
Ordered phase and scaling in $Z_n$ models and the three-state antiferromagnetic 
Potts model in three dimensions, 
Phys. Rev.B 61, 3430--3434 (2000).
 


%
%









 
 
 
 
 
 
 

%
 




 

\bibitem{Po} R.~B.~Potts: Some generalized order-disorder transformations,
Math. Proc. Cambridge Phil. Soc. 48, 106--109 (1952). 

%



\bibitem{Ruth} W.M.~Ruszel: Gibbs and non-Gibbs aspects of continuous spin models, Groningen thesis, (2010). 

\bibitem{TUM} N.~Todoroki, Y.~Ueno and S.~Miyashita: Ordered phase and phase transitions in the three-dimensional generalized six-state clock model, Phys. Rev. B 66, 214405 (2002).

\bibitem{SI93} P.D.~Scholten and L.J.~Irakliotis: Critical behavior of the q-state clock model in three dimensions,  Phys. Rev. B. 48, 1291--1294 (1993).

\bibitem{TO} Y. Tomita and Y. Okabe: Probability-changing cluster algorithm for two-dimensional XY and clock models, Phys. Rev. B65, 184405 (2002).

\bibitem{Ue} Y. Ueno: Description of ordering and phase transition in terms of 
local connectivity: Proof of a novel type of percolated state in the general clock model, J. Stat. Phys. 80, 843--870 (1995).

\bibitem{UK} Y.~Ueno and K. Kasono:
Incompletely ordered phases and phase transitions ih the three-dimensional general clock model, Phys.Rev. B 48, 16471 (1993).

\bibitem{Ver} E.A. Verbitskiy: Variational principle for fuzzy Gibbs measures, Moscow Math. J. 10, 811-829 (2010). 



\end{thebibliography}
\end{document}